\documentstyle[12pt]{article}

\begin{document}
\hfill\vbox{
\hbox{OHSTPY-HEP-T-96-017}
\hbox{hep-th/9609047}
\hbox{September 1996} }\par
\thispagestyle{empty}

\vspace{.5in}

\begin{center}
{\Large \bf Renormalization Effects in a Dilute Bose Gas}

\vspace{.5in}

Eric Braaten and Agustin Nieto \\

{\it Department of Physics, The Ohio State University, Columbus, OH 43210}

\end{center}

\vspace{.5in}

\begin{abstract}
The low-density expansion for a homogeneous
interacting Bose gas at zero temperature
can be formulated as an expansion in powers of $\sqrt{\rho a^3}$,
where $\rho$ is the number density and $a$ is the S-wave scattering length.
Logarithms of $\rho a^3$ appear in the coefficients of the expansion.
We show that these logarithms are determined by the renormalization
properties of the effective field theory that describes the scattering
of atoms at zero density.  The leading logarithm is determined
by the renormalization of the pointlike $3 \to 3$ scattering amplitude.
\end{abstract}

\newpage

The successful achievement of Bose-Einstein condensation of atomic
gases in magnetic traps \cite{trap} has created an explosion of interest
in Bose gases of atoms.
While a qualitative description of the condensation can be obtained using
mean field methods \cite{Baym-Pethick},
a more quantitative treatment requires
including corrections from fluctuations around the mean field.
The relative magnitude
of these corrections grows with the number density of the atoms.
They will therefore become more important as higher
trap densities are achieved.

In order to develop a deeper understanding of the fluctuations, it is
worthwhile to go back to the simpler problem of a homogeneous gas
of interacting bosons at zero temperature.
This problem was studied intensively in the 1950's
\cite{Lee-Yang,Wu}.
A simple review was given by Yang in 1960 \cite{Yang}.
The properties of the system can be calculated as an expansion in powers of
$\sqrt{\rho a^3}$, where $\rho$ is the number density of atoms
and $a$ is their S-wave  scattering length.  For example,
the expansion for the energy density has the form
\begin{equation}
{\cal E} \;=\;  {2 \pi \rho^2 a \over m}
\left\{ 1 \;+\; {128 \over 15 \sqrt{\pi}} \sqrt{\rho a^3}
	\;+\;  \left[ {8 \over 3} ( 4 \pi - 3 \sqrt{3} )
		\log (\rho a^3) + \kappa \right] \rho a^3
	\;+\; \ldots \right\}  \;,
\label{E-Yang}
\end{equation}
where we have set $\hbar = 1$.
The coefficient of $\sqrt{\rho a^3}$ was first obtained by Lee and Yang
for a hard sphere gas \cite{Lee-Yang}.
The coefficient of $\rho a^3 \log(\rho a^3)$
was calculated by Wu, by Hugenholtz and Pines,
and by Sawada \cite{Wu}.  The correction  $\kappa \rho a^3$
is the first term in the expansion that is sensitive to atomic parameters
other than the scattering length. We have recently succeeded in
calculating the constant $\kappa$ \cite{Braaten-Nieto}.

In this Letter, we use a minimal subtraction renormalization
scheme to deduce the general structure of the low-density expansion.
We show that logarithms of $\rho a^3$ are related to the
renormalization of the effective field theory that describes the
scattering of atoms in the vacuum.   In particular, the $\log(\rho a^3)$
term in (\ref{E-Yang}) is related to the
renormalization of the $3 \to 3$ scattering amplitude.
We reproduce the leading logarithms in previous calculations
using simple renormalization group methods.  Our approach can also be used to
determine the logarithms
that appear at higher orders in the low-density expansion.

Our starting point is an effective field theory \cite{Georgi}
that describes atoms
with momenta much lower than their inverse size, which is on the
order of the  Bohr radius $a_0$.  Since the range of the
interaction potential between 2 or more atoms is also on the order
of $a_0$, the interactions appear pointlike on the scale of the
de Broglie wavelengths of the atoms.  The atoms can therefore
be described by a field theory with a hamiltonian density that is
a local function of the field:
\begin{equation}
{\cal H} \;=\;
- {1 \over 2 m} \psi^\dagger \nabla^2 \psi
\;+\; {1 \over 4} g (\psi^\dagger \psi)^2
\;+\; {1 \over 36} g_3 (\psi^\dagger \psi)^3
\;+\; \ldots,
\label{H-eff}
\end{equation}
For simplicity, we have assumed that the atoms have spin 0,
so that they can be represented by a single complex field $\psi$.
The $(\psi^\dagger \psi)^2$ term represents $2\to 2$
scattering through an S-wave interaction
with scattering length $a$ given by $g = 8 \pi a / m$, while the term
$(\psi^\dagger \psi)^3$ represents $3 \to 3$ scattering.
By adding additional terms that are higher order in the
derivatives or in the number of fields, one can describe $n \to n$
scattering of atoms in the vacuum with whatever accuracy is desired.
In principle, the coefficients of these terms can be calculated
from the $n$-body potentials that describe interatomic interactions.
In the absence of such
calculations, they can be taken as phenomenological parameters.

By treating the interaction terms as perturbations,
we can calculate the amplitudes for
scattering of atoms with momenta on the order of
$p$ as an expansion  in powers of $p a_0$.  This expansion is
complicated by the presence of ultraviolet divergences.
For example, the amplitude for the scattering of two atoms
with momenta ${\bf p}_1$ and  ${\bf p}_2$,
including the first perturbative correction, is
\begin{equation}
g \left[ 1 \;-\; {m g \over 2} \int {d^3k \over (2 \pi)^3}
	{1 \over k^2 - ({\bf p}_1 + {\bf p}_2) \cdot {\bf k}
		+ {\bf p}_1 \cdot {\bf p}_2 - i \epsilon} \right] .
\label{g-1loop}
\end{equation}
The integral, which is ultraviolet divergent, can be
regularized by imposing a cutoff $|{\bf k}| < \Lambda$.
The linear divergence can then be cancelled by adding a counterterm
proportional to $m g^2 \Lambda (\psi^\dagger \psi)^2$ to the
effective hamiltonian (\ref{H-eff}).  The resulting expression
for the scattering amplitude is rather complicated, as it
includes terms that are suppressed by powers of
$p_1/\Lambda$ and $p_2/\Lambda$.  A simple analytic result is obtained
only in the limit $\Lambda \to \infty$.

A power ultraviolet divergence, such as the linear divergence in
(\ref{g-1loop}), indicates extreme sensitivity
to short-distance atomic physics that is not accurately described
by the effective hamiltonian (\ref{H-eff}).  A simple momentum cutoff
is not an accurate model for the way atomic physics cuts off the momentum
integrals.
There is an alternative cutoff procedure, called ``minimal subtraction'',
which is no more accurate a model for the cutoff, but provides
an equally accurate description of the long-distance physics and
has the virtue of simplicity.
In minimal subtraction, linear, quadratic, and other power
ultraviolet divergences are
removed as part of the regularization scheme by subtracting
the appropriate power of $k$ from the momentum space integrand.
In the case of the amplitude (\ref{g-1loop}),
$1/k^2$ is subtracted from the integrand.
The justification for this procedure is that
the terms that are subtracted
are dominated by short distances and can be cancelled by
counterterms in the Hamiltonian.
In minimal subtraction, logarithmic ultraviolet divergences
are treated differently from power divergences.
This is reasonable, because logarithmic ultraviolet divergences
represent real physical effects, while power ultraviolet divergences
are simply artifacts of the regularization procedure.  This
difference is reflected in the fact that the coefficient of a power
divergence $\Lambda^p$ depends on the regularization prescription,
while the coefficient of $\log(\Lambda)$ does not.  The reason for this
is that the logarithm of $\Lambda$ must match onto the logarithm
of some physical momentum scale, and therefore its coefficient
has a real physical meaning.
We regularize logarithmic ultraviolet divergences by imposing a
cutoff $|{\bf k}| < \Lambda$ on loop integrals.
After using renormalization to remove divergences from subdiagrams,
we isolate the divergent terms proportional to $\log(\Lambda)$ and then take
the limit $\Lambda \to \infty$ in the remainder.
The cutoff $\Lambda$ is called the ``renormalization scale''.
With minimal subtraction, all power divergences and those parts of
logarithmic divergences that arise from  momenta greater than
the renormalization scale $\Lambda$ are absorbed into the coupling
constants in the effective hamiltonian.

The advantage of minimal subtraction is
that it makes it much easier to disentangle the
effects of different momentum scales in multiloop diagrams.
With a conventional momentum cutoff,  a diagram can be an
extremely complicated function of the cutoff $\Lambda$, the external
momenta, and the momentum scales that can be formed from the
parameters in the hamiltonian.  With minimal subtraction,
the possible dependence of a diagram on the cutoff
is greatly simplified. The dependence can only be polynomial in
$\log(\Lambda)$, with the logarithms arising from
logarithmically ultraviolet divergent subdiagrams.
This makes it much easier to analyze the divergences in a multiloop diagram.
The relative simplicity of minimal subtraction is illustrated by the
fact that it gives a simple expression for the scattering amplitude
(\ref{g-1loop}) that is independent of the renormalization scale $\Lambda$:
$g [1 + i m g |{\bf p}_1 - {\bf p}_2|/(16 \pi)]$.
The renormalized parameter $g$ is independent of $\Lambda$,
and satisfies a trivial renormalization group equation:
$\Lambda (d/d \Lambda) g = 0$.
With a conventional momentum cutoff $\Lambda$, the amplitude
(\ref{g-1loop}) is a complicated function of
${\bf p}_1$, ${\bf p}_2$, and  $\Lambda$.
For $\Lambda \gg |{\bf p}_1|, |{\bf p}_2|$, it reduces to
$g(\Lambda) [1 - i m g |{\bf p}_1 - {\bf p}_2|/(8 \pi)]$,
where $g(\Lambda)$ is a renormalized coupling constant that satisfies
$\Lambda (d/d \Lambda) g(\Lambda) = - m g^2 \Lambda/(4 \pi^2)$.
The running of $g(\Lambda)$ is generated by a power ultraviolet
divergence and therefore has no real physical significance.
The scale-invariant parameter $g$ defined by minimal subtraction
provides an equally accurate description of the long-distance physics.

In the vacuum, the simplest quantity in which
logarithmic ultraviolet divergences appear is the $3 \to 3$
scattering amplitude.  There is a tree-level contribution from the
$(\psi^\dagger \psi)^3$ term in the
effective hamiltonian, but there are also
additional contributions that involve successive $2 \to 2$ scatterings.
They include the 2-loop diagrams shown in Figure 1, which involve
4 successive $2 \to 2$ scatterings.  These diagrams are logarithmically
ultraviolet divergent.  Removing the linear ultraviolet divergence
from a subdiagram of the first diagram in Figure 1 by a subtraction
in the integrand
and then imposing a cutoff $\Lambda$, we find that the logarithmically
divergent term is
$ - 3 (4 \pi - 3 \sqrt{3}) m^3 g^4 \log (\Lambda) / (32 \pi^3)$.

The renormalization scale $\Lambda$ represents an arbitrary separation
between short-distance effects,
which are taken into account through the parameters in the effective
hamiltonian (\ref{H-eff}), and long-distance effects, which are calculated
using the effective theory.
Physical quantities, such as the $3 \to 3$ scattering amplitude,
should therefore be completely independent of $\Lambda$.
The explicit $\Lambda$-dependence from the two-loop diagrams must
therefore be cancelled by implicit $\Lambda$-dependence from the
coefficient $g_3$ in the tree-level contribution.
This statement can be expressed as a ``renormalization group equation'':
\begin{equation}
\Lambda {d \ \over d \Lambda} g_3(\Lambda)
\;=\; {3 \over 32 \pi^3} ( 4 \pi - 3 \sqrt{3} ) m^3 g^4 \;.
\label{rge}
\end{equation}
It tells us that the parameter $g_3$ is really a ``running coupling constant''
that increases logarithmically as the momentum scale is increased.
The renormalization scale $\Lambda$ can be interpreted as the inverse of
the spacial resolution.  As this resolution is decreased, we resolve
part of the ``pointlike'' $3 \to 3$ scattering amplitude into the successive
$2 \to 2$ scatterings represented by the diagrams in Figure 1.
The contributions from the two diagrams have opposite signs
and the net effect is that the coupling constant $g_3$ decreases
as $\Lambda$ decreases.

If the running coupling constant $g_3$ is determined at the scale $1/a_0$ of
atomic structure, it can be calculated at a lower momentum scale $\Lambda$
by solving the renormalization group equation (\ref{rge}):
\begin{equation}
g_3(\Lambda) \;=\; g_3(1/a_0)
\;-\; {3 \over 32 \pi^3} ( 4 \pi - 3 \sqrt{3} ) m^3 g^4
	\log \left({1 \over \Lambda a_0} \right)  \;.
\label{rgsol}
\end{equation}
Regardless of the sign of $g_3(1/a_0)$, $g_3(\Lambda)$ eventually
turns negative for sufficiently small $\Lambda$.
In describing the scattering of atoms with momenta on the order of $p$,
it is appropriate to choose the renormalization scale $\Lambda$ to be
of order $p$.  The coefficients in the perturbation expansion
can include logarithms of the form $\log(\Lambda/p)$,
which are generated by logarithmically divergent subdiagrams.
By choosing $\Lambda$ to be of order $p$,
such large logarithms are removed from the coefficients
and absorbed into the parameters of the effective hamiltonian.

We now consider the energy density $\cal E$ of the Bose gas with number
density $\rho$.  In order for the system to have a homogeneous
ground state that is stable, or at least metastable, the
scattering length $a$ must be positive.
Neglecting for the moment the effects of fluctuations,
the field $\psi$ develops a vacuum expectation value $\sqrt{\rho}$.
The energy density at tree level is
\begin{equation}
{\cal E}_0 \;=\; {1 \over 4} g \rho^2
	\;+\; {1 \over 36} g_3(\Lambda) \rho^3  \;+\; \ldots \;.
\label{E0}
\end{equation}
Setting $g = 8 \pi a/m$, where $a$ is the S-wave scattering length,
the first term above reproduces the leading term in (\ref{E-Yang}).
The 1-loop contribution is the sum of the zero-point energies of the
Bogoliubov modes, with power ultraviolet divergences removed
by subtractions in the integrand:
\begin{equation}
{\cal E}_1 \;=\; {1 \over 2} \int {d^3 k \over (2 \pi)^3}
\left[ \epsilon(k) \;-\; {k^2 \over 2m} \left( 1 + {g m \rho \over k^2}
			- {g^2 m^2 \rho^2 \over 2 k^4} \right) \right] \;,
\label{E1}
\end{equation}
where $\epsilon(k) = k \sqrt{k^2 + 2 m g \rho} / (2m)$.
This integral reproduces the first correction term in
(\ref{E-Yang}).

The correction of order $\rho a^3$ in (\ref{E-Yang}) requires
the calculation of two-loop diagrams.  However the term
proportional to $\rho a^3 \log(\rho a^3)$ can be obtained without
any further calculation.  The reason is that
this term is related to the renormalization of the amplitude for
$3 \to 3$ scattering in the vacuum.
The two-loop diagrams for the energy density contain logarithmic
ultraviolet divergences.
 From the expression for the Bogoliubov energy, we see that
the momentum scale associated with the quasiparticles modes
is $\sqrt{2 m g \rho}$.  The logarithmic ultraviolet divergences
from the two-loop diagrams will therefore be proportional to
$\log(\Lambda / \sqrt{2 m g \rho})$.  Large logarithms such
as this in the coefficients in the perturbation expansion
can be avoided by choosing the renormalization scale $\Lambda$
to be on the order of $\sqrt{2 m g \rho}$.  With this choice of the
renormalization scale, all such logarithms are absorbed into
the parameters in the effective hamiltonian.
Substituting $\Lambda = \sqrt{16 \pi a \rho}$ in (\ref{E0}) and using the
expression for $g_3(\Lambda)$ in (\ref{rgsol}),
we reproduce the term containing the logarithm in (\ref{E-Yang}).
We have determined the constant $\kappa$ under the logarithm
by calculating the 2-loop Feynman diagrams
for the energy density explicitly.
The details of the calculation will be reported elsewhere
\cite{Braaten-Nieto}. As noted previously \cite{Wu}, the $\rho a^3$
term is the first term in the low density expansion (\ref{E-Yang})
that is sensitive to atomic physics parameters other than the
scattering length $a$. The only additional parameter that enters
at this order is the pointlike $3 \to 3$ scattering amplitude $g_3$.

The renormalization group together with minimal subtraction
can also be used to determine the leading logarithms
in the low density expansions for other quantities.
Corrections to the sound velocity  have been calculated by Beliaev
\cite{Beliaev}, including the $\sqrt{\rho a^3}$ term and the
$\rho a^3 \log(\rho a^3)$ term.  The logarithm can be
obtained by calculating the logarithmic ultraviolet divergences
in the two-loop corrections to the propagator.  Alternatively,
it can be obtained trivially using the methods described above.
Including the correction from the $(\psi^\dagger \psi)^3$ term
in (\ref{H-eff}), the sound velocity at tree level is
$v^2 = g \rho/(2 m) + g_3(\Lambda)\rho^2/3$.
Setting $\Lambda = \sqrt{16 \pi a \rho}$ and using the expression
(\ref{rgsol}) for $g_3(\Lambda)$, we reproduce the
$\rho a^3 \log(\rho a^3)$ correction calculated by Beliaev.

These methods can also be used to determine the coefficients of
the logarithms that appear at higher orders in the low-density
expansion for the energy density.
For example, there is a $(\rho a^3)^{3/2} \log(\rho a^3)$
correction to the energy density which arises from
logarithmically divergent two-loop subdiagrams in 3--loop diagrams.
This terms can be determined easily by taking into account
the $(\psi^\dagger \psi)^3$ term in the Bogoliubov energy:
$\epsilon^2(k) = k^2 (k^2 + 2 m g \rho)
	/ (4 m^2) + g_3(\Lambda) \rho^2 k^2/3$.
Inserting this into the 1-loop expression (\ref{E1}) for the energy density,
expanding to first order in $g_3$, and using the expression
(\ref{rgsol}) for $g_3(\sqrt{16 \pi a \rho})$, we obtain the
$(\rho a^3)^{3/2} \log(\rho a^3)$ correction.

Thus far we have only considered logarithms in the low density
expansion that are related to the renormalization of the $3 \to 3$
scattering amplitude.  At higher orders in the low density expansion,
there are also logarithms that are related to
the renormalization of other terms in the effective lagrangian, such as
the term $g_4 (\psi^\dagger \psi)^4$
which describes $4 \to 4$ scattering through a point-like interaction.
The $4 \to 4$ scattering amplitude includes logarithmically ultraviolet
divergent corrections from 4-loop diagrams that
involve 7 successive $2 \to 2$ scatterings and also from 2-loop
diagrams that involve  three $2 \to 2$ scatterings
and a  $3 \to 3$ scattering.  The explicit $\Lambda$--dependence from
these loop diagrams must be cancelled by the implicit $\Lambda$--dependence
from the pointlike $4 \to 4$ scattering amplitude $g_4(\Lambda)$.
As a consequence, the renormalization
group equation analogous to (\ref{rge}) for $(\Lambda d/d \Lambda) g_4$
includes terms on the right side that are proportional to $m^6 g^7$
and $m^3 g^3 g_3$.  The solution for $g_4(\Lambda)$ analogous to
(\ref{rgsol}) includes a term proportional to
$m^6 g^7 \log^2(\Lambda a_0)$.  Choosing $\Lambda = \sqrt{16 \pi a \rho}$
to avoid large logarithms from loop diagrams, we find that the term
$g_4 \rho^4$ in the mean-field expression for the energy density
gives rise to a correction to (\ref{E-Yang}) that is proportional to
$(\rho a^3)^2 \log^2(\rho a^3)$.

A renormalization group analysis of the dilute Bose gas at nonzero
temperature was recently carried out in Ref. \cite{Bijlsma-Stoof}.
The authors derived renormalization group equations for the
chemical potential $\mu$  and for the coupling constant $V_0 = 2 g$
using a conventional momentum cutoff.  If they had used
minimal subtraction, their analysis would not
have been modified dramatically.  Their equations for $d \mu/d l$
would remain unchanged.  In their equations for $d V_0 / d l$,
there would be an additional term $V_0^2 m \Lambda/(2 \pi^2)$
on the right side which cancels the leading power of $\Lambda$
in the equation at $T=0$.  The renormalization group trajectories
for these two renormalization schemes would differ significantly
only near the initial cutoff, where both calculations
would be dominated by cutoff artifacts.  They would be essentially
identical near the critical point for Bose condensation.
The logarithmic evolution of the coupling constant $g_3$ was not
seen in the analysis of Ref. \cite{Bijlsma-Stoof}, because they
considered renormalization effects from one-loop diagrams only.

In this Letter, we have shown how the structure of the low-density
expansion for a Bose gas is determined by the renormalization
properties of the effective hamiltonian that describes the
scattering of atoms at zero density.  The low density expansion
for the energy density has the general form
\begin{equation}
{\cal E} \;=\;  {\rho^2 a \over m}
\sum_{n=0}^\infty \sum_{l=0}^{l_n}  C_{nl} \;
	(\rho a^3)^{n/2} \log^l(\rho a^3) \;.
\label{E-gen}
\end{equation}
The maximum power of the logarithm has been determined to be
$l_n=0,0,1,1,2$ for $n=0,1,2,3,4$, respectively.
The dimensionless coefficients $C_{nl}$ are polynomials in the
generalized coupling constants of higher order terms in the
effective hamiltonian (\ref{H-eff}), with only a finite number of
these coupling constants appearing at any given order in
$\sqrt{\rho a^3}$.  The coupling constant $g_3$ first appears
at order $\rho a^3$.  Additional coupling constants
enter at order  $(\rho a^3)^{3/2}$.
The general structure in (\ref{E-gen}) follows automatically
from the renormalization group together with the
minimal subtraction renormalization scheme.
This powerful method should also be useful
for analyzing the corrections from fluctuations around the mean field
for atomic gases in magnetic traps.

This work was supported in part by the U.~S. Department of Energy,
Division of High Energy Physics, under Grant DE-FG02-91-ER40690.

\vspace{.5in}

\noindent
{\Large \bf Figure Caption}

\noindent
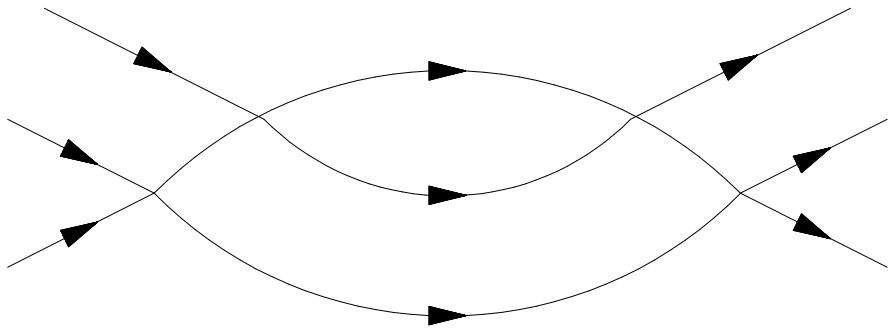
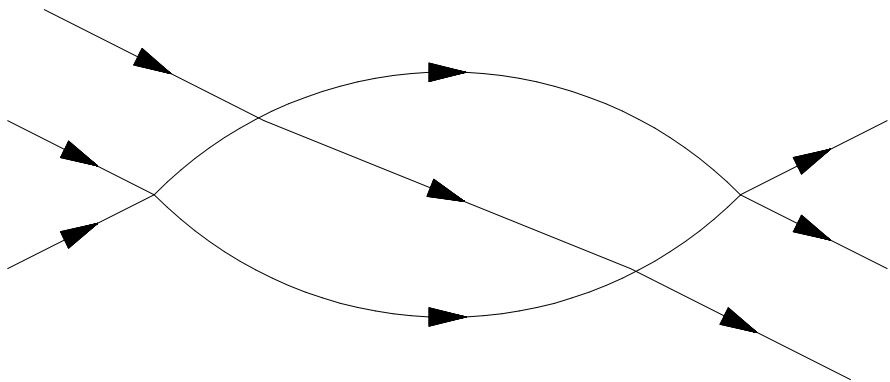
Figure 1.  Two-loop diagrams that give logarithmically ultraviolet
divergent contributions to the $3 \to 3$ scattering of atoms.

\end{document}